\documentclass[twocolumn,aps,reprint, nofootinbib]{revtex4}
\usepackage{graphicx}
\usepackage{epsfig,amsmath}
\usepackage{amssymb}
\usepackage{rotate}
\usepackage{color}
\usepackage[ps2pdf=true]{hyperref}
\usepackage{bm}

\newcommand{\be}{\begin{eqnarray}}
\newcommand{\ee}{\end{eqnarray}}

\newcommand{\Omegam}{{\Omega_\mathrm{m}}}
\newcommand{\Omegal}{{\Omega_\Lambda}}

\newcommand{\e}{{\mathrm{e}}}
\newcommand{\ii}{{\mathrm{i}}}

\newcommand{\dd}{{\rm d}}

\newcommand{\eff}{{\rm eff}}
\newcommand{\w}{{w_{\eff}}}

\newcommand{\bfth}{{\boldsymbol{\theta}}}

\newcommand{\Dirac}{\delta_{\rm D}}

\newcommand{\hkappa}{\hat{\kappa}_{_<}}
\newcommand{\hgamma}{\hat{\gamma}}
\newcommand{\hsigma}{\hat{\sigma}}

\newcommand{\mD}{{\mathcal{D}_0}}
\newcommand{\tU}{\tilde{U}}
\newcommand{\tQ}{\tilde{Q}}
\newcommand{\tV}{\tilde{V}}
\newcommand{\vark}{\vartheta_{\hkappa^{lin}}}
\newcommand{\Map}{\hat{M}_{ap}}
\newcommand{\Mapg}{\hat{M}_{ap}^g}

\newcommand{\varphik}{\varphi^{\kappa}}
\newcommand{\varphig}{\varphi^{\gamma}}
\newcommand{\varphir}{\varphi^{g}}

\newcommand{\pr}[1]{{\textcolor{black}{\textit{} #1}}}

\begin{document}
\title{Large deviations theory approach to cosmic shear calculations: the one-point aperture mass}

\author{Paulo Reimberg}
\email{reimberg@iap.fr}
\affiliation{Sorbonne Universit\'es, UPMC Univ Paris 6 et CNRS, UMR 7095, Institut d'Astrophysique de Paris, 98 bis bd Arago, 75014 Paris, France}
\author{Francis Bernardeau}
\affiliation{Sorbonne Universit\'es, UPMC Univ Paris 6 et CNRS, UMR 7095, Institut d'Astrophysique de Paris, 98 bis bd Arago, 75014 Paris, France}
\affiliation{CEA - CNRS, UMR 3681, Institut de Physique Th\'eorique, F-91191 Gif-sur-Yvette, France}

\begin{abstract}
This paper presents a general formalism that allows the derivation of the cumulant generating function and one-point Probability Distribution Function (PDF) of the aperture mass ($\Map$), a common observable for cosmic shear observations. Our formalism is based on the Large Deviation Principle (LDP) applied, in such cosmological context, to an arbitrary set of densities in concentric cells. We show here that the LDP can indeed be used for a much larger family of observables than previously envisioned, such as those built from continuous and nonlinear functionals of the density profiles. 
The general expression of the observable aperture mass depends on reduced shear profile making it a rather involved function of the projected density field. Because of this difficulty, an approximation that is commonly employed consists in replacing the reduced shear by the shear in such a construction neglecting therefore non-linear effects.
We were precisely able to quantify how this approximation affects the $\Map$
statistical properties. In particular we derive  the corrective term for the skewness of the $\Map$ and reconstruct its one-point PDF.
\end{abstract}


\maketitle


\section{Introduction}

Gravitational lensing has been shown to be a very efficient way of exploring the properties of the mass distribution at large scale. It indeed
provides information about the gravitational potential that light rays go through, from sources to observer. Although it is fair to say that the most spectacular consequences of such phenomena are the strong lensing effects, with the occurrences of multiple images and large arc-like deformation of images of background objects, the most fruitful regime  in the context of cosmological observation is the one of the cosmic shear: weak lensing effects are indeed ubiquitous and are induced by the large-scale structure of the universe as a whole. 
The first evidences that large-scale structure can coherently affect the shape of background galaxies have been presented in year 2000 in a series of compelling results, \cite{2000Natur.405..143W,2000A&A...358...30V,2000MNRAS.318..625B}. The evidence is based on the fact that such deformations are expected to obey a specific geometrical property, namely the absence of parity-odd contributions (i.e. negligible $B$-modes).
These results opened the way to a systematic use of such observations to map the large-scale structure of the universe and explore its statistical properties. To be more specific, in such a regime (in the absence of critical region), the information provided by cosmic shear observations are encoded in the elements of a deformation matrix, the convergence and shear fields, that describe the magnification and deformation of the shape of light beams. The reconstruction of such maps from background galaxy shapes would provide in principle direct information about the projected mass \cite{2003astro.ph..5089V,  munshi2008cosmology, kilbinger2015cosmology}.  This is key to a large part of the core programs of projects such as the CFHTLenS\footnote{\texttt{www.cfhtlens.org}}, the Dark Energy Survey (DES)\footnote{\texttt{www.darkenergysurvey.org}}, LSST\footnote{\texttt{www.lsst.org}}, and Euclid\footnote{\texttt{www.euclid-ec.org}}.

Cosmic shear observations are usually exploited with the help of the shear two-point correlation function (or equivalently its power spectrum). This is what the approach which is usually adopted such as in 
\cite{2008A&A...479....9F} for the CFHTLenS survey
and in 
\cite{2016PhRvD..94b2002B} for the DES survey. There exist however alternative approaches that give complementary information. This is in particular the case of the convergence one-point Probability Distribution Function (PDF) or rather the PDF of the aperture mass (which is a specific filtering of the convergence map we define below) and its first few moments as exploited in 
\cite{2014MNRAS.441.2725F}
for the CFHTLS or in
\cite{2016MNRAS.463.3653K}
for the DES survey.

This paper is in this line of investigation. It does not aim at quantifying the efficiency of such measurements in constraining the cosmological models but at showing the one-point PDF of the aperture mass can be computed from first principles in a given cosmological context. Such a computation makes use of the Large Deviations Principle (LDP), \pr{that is the central definition on the theory of large deviations, and} that we will describe more specifically in the following. 
In broad terms, the theory of large deviations \cite{varadhan, der_hollander, ellis} is a branch of probability theory that  deals with rate at which probabilities of certain events decay as a natural parameter of the problem varies \cite{varadhan_lec_notes}, and is applied in variety of domains in mathematics and theoretical physics, specially in statistical physics both for equilibrium and non-equilibrium systems (see for instance \cite{2009PhR...478....1T} for a review paper on the subject, or App. \ref{Large_dev} for a short introduction). The application of the LDP to Large Scale Structure cosmology has been formalized in the last few years \cite{2002A&A...382..412V, bernardeau2014statistics, Bernardeau:2015khs, uhlemann2016back} and will be employed here in the context of cosmic shear observations.  This is not the first time that arguments related to the theory of large deviations is used in this context, and we highlight \cite{bernardeau2000construction, munshi2004weak, valageas2004analytical, barber2004linear} in particular.

In this paper we focus on aperture mass statistics and extend the results that had been obtained in two ways: the aperture mass is defined with the help of more realistic filters, and we show that one can take into account the fact that only the reduced shear can be measured, not the shear itself. Cosmic shear observations are indeed based on the measurements of background galaxy shapes, more specifically on the amplitude and direction of their deformation, and what we have access to are ratios of the deformation matrix elements, i.e; the reduced shear which is then the ratio (shear)/(1-convergence). Moreover, from such observation one can easily build the Laplacian of the associated (reduced) convergence field.  But it is not possible to unambiguously recover the convergence field itself. For circular symmetric filters, the convergence can only be recovered in compensated filters (which filter out constant fields and fields with constant gradients) which can be viewed as projected mass maps in aperture\footnote{Conversely it can be shown that such Map fields can be written in terms of integrals of well chosen components the reduced shear field.}, that is \emph{aperture mass} (Map) \cite{kaiser1995nonlinear, schneider1996detection}. 

In general observations should then be viewed as non local, non linear transformation of the convergence field. In the literature the reduced shear is usually replaced by the shear itself arguing that the convergence ought to be generally small for cosmic shear observations. It dramatically simplifies the problem as the observed Map is made a linear transformation of the projected mass field. This is however an a priori unjustified simplification in the context of the theory of large deviations. It can be shown that the statistical properties of quantities that respect certain symmetries (which here will be the circular symmetry) can be computed in the small variance limit even though some events strongly depart form the variance. What the LDP can account for is precisely the impact  of excursions of large values of the convergence. 

The general techniques for computing one-point PDF of densities filtered 
with top-hat filters has been developed in a long series of papers \cite{1994A&A...291..697B, 1994ApJ...433....1B, Bernardeau:1994hn, bernardeau2014statistics}. The derivation of exact results for continuous filters escaped however this formalism\footnote{The derivation of the skewness when gaussian filtering is employed, for example, was a remarkable tour de force \cite{1993ApJ...412L...9J}.}. The central ingredient of the developments presented here is the construction of general symmetric filtering of the density field as the continuous limit of a weighted composition of concentric top-hat filters, as introduced in \cite{Bernardeau:2015khs}. The second main point of this paper is to show that the LDP can be applied to  a non-linear functional of the density profile and namely to the one-point PDF of the observed Map. Again this construction will make use of the general formulation developed in \cite{Bernardeau:2015khs}. Our work therefore extends previous results in two fundamental ways.

The application of the LDP is based on a number of ingredients we will detail in section II. The line of reasoning we follow (similar to \cite{bernardeau2000construction}) is based on the fact that cosmic shear observations are akin to observation in long cylindrical cells. In the following we will simply assume that the fluctuations along the radial direction have simply been integrated out. The LDP is then based on the assumption that the leading configurations in the initial field that lead, after nonlinear evolution, to a  given circular constraint obey the same circular symmetry. Under this assumption (and before shell crossing) it is then possible to map the initial configuration to the final one.
The setting of the LDP is then based on the following ingredients:
\begin{itemize}
\item One should first define the rate function of the variables that define  the initial field configuration. We will assume here Gaussian initial condition to define their covariance matrix.
\item One should then specify the mapping between the initial field configuration (mass profile) and the final mass profile. It will make use of the 2D cylindrical collapse (or rather one approximation of it).
\item We are then in position to write the observable -- say the Map defined with a specific filter --  as a functional form of the final and therefore initial mass profile.
\end{itemize}

The rate function of the Map variable is then obtained through a minimization procedure we detail in section III.
The results are discussed in section IV and in particular we comment on the impact of the differences between Map and observed Map
statistics at the level of the rate function, the cumulant generating function and the resulting one-point PDF.

\section{The LDP applied to the convergence field}

The aim of this section is to introduce the necessary ingredients required for the implementation of the LDP. We first review 
the nature of the observable we are interested in.

\subsection{The convergence}

\begin{figure}[!ht]
\centering
\includegraphics[width=7.5cm]{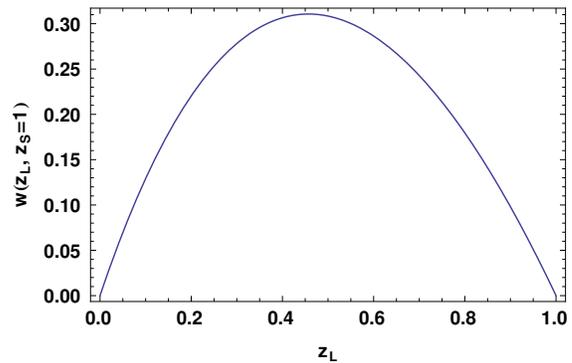}
\caption{Projection factor  $w(z_{_L}, z_{_S})$ defined in Eq. \eqref{w_definition}  in Einstein- de Sitter universe for $z_{_S}=1$. The effective value $\w$ is the particular values of $w(z_{_L}, z_{_S})$ at $z_{_L} = z_{_L}^*$ corresponding to the redshift of the effective lens plan.}
\label{fproj}
\end{figure}

\pr{Lensing effects are generally classified in strong and weak. On the first case arcs and multiple images are observable as consequence of caustics and cusps on the observers past-light cone due to gravitation. Weak gravitational lensing is associated to smooth deformations on the observers past-light cone and the effects are less dramatic. Gravitational lensing on distant galaxies can be observed through  the distortion of shape and size of the light sources parametrized by the \emph{shear} $\gamma$ and \emph{convergence} $\kappa$, respectively. } For scalar, linear perturbations of the FLRW spacetime, convergence and shear can be expressed as derivatives of a \emph{lens potential} (see App. \ref{app_lens}).  The convergence, in particular, is given by $2 \kappa = (\triangledown_1 \triangledown_1 +  \triangledown_2 \triangledown_2) \phi_L$, where $\triangledown_{1, 2}$ are derivatives on a spacelike plan perpendicular to the direction of propagation of the light beam. \pr{If we assume Born approximation, neglect lens-lens couplings and extrapolate the 2D Laplacian that appears on the definition of $\kappa$ to the 3D Laplacian, we can use the Poisson's equation to link convergence and matter density fluctuations as (see for instance \cite{munshi2008cosmology})}:
\begin{equation}
\kappa(\bfth) =  \int_0^{\chi_{_S}} \dd \chi_{_L} w(\chi_{_S}, \chi_{_L}) \delta(\mD(\chi_{_L}) \bfth, \chi_{_L}) \, .
\end{equation}
Here $\chi_{_S}$ is the comoving distance to the sources, and
\begin{equation}
\label{w_definition}
w(\chi_{_S}, \chi_{_L}) := \frac{3}{2} \Omegam \frac{H_0^2}{c^2} \frac{\mD(\chi_{_S} - \chi_{_L}) \mD(\chi_{_L})}{a(\chi_{_L}) \mD(\chi_{_S})}
\end{equation} 
with $\mD$ is defined in Eq. \eqref{ang_dist}. We observe that $w(\chi_{_S}, \chi_{_L})$ is
 a positive function, and therefore by the integral Mean-Value theorem, there exists $\chi_{_L}^*$, $0 < \chi_{_L}^* < \chi_{_S}$, such that

\begin{equation}
\label{kappa_tvm}
\kappa(\bfth) =  \w(\chi_{_S}) \, \delta_{2D}(\mD(\chi_{_S}) \bfth, \chi_{_S})
\end{equation}
\pr{we will not keep the dependence on $\chi_{_S}$ explicitly in $\kappa(\bfth)$  to make the notation shorter. We define $\w(\chi_{_S})$ explicitly as:}
\begin{equation}
\w(\chi_{_S}) := \frac{3}{2} \Omegam \frac{H_0^2}{c^2} \frac{\mD(\chi_{_S} - \chi_{_L}^*) \mD(\chi_{_L}^*)}{ a(\chi_{_L}^*)  \mD(\chi_{_S})} \, .
\end{equation}
\pr{This corresponds effectively to existence os a \emph{lens plan} at $\chi_{_L}^*$ where all the lensing mass would be concentrated. We could have defined a \emph{projected density contrast} instead, as in \cite{bernardeau2000construction}, but the assumption of a lens plan does not weaken our present argument. In Einstein-de Sitter universe, the possible values of $ \w( z_{_S})$ are shown in Fig. \ref{fproj} as function of the possible $z_{_L}^*$s.}

\pr{The quantity $\delta_{2D}(\mD(\chi_{_S}) \bfth, \chi_{_S})$ is given by:}
\begin{equation}
\label{delta_projected}
\delta_{2D}(\mD(\chi_{_S}) \bfth, \chi_{_S}) = \int_0^{\chi_{_S}} \dd \chi'  \delta(\mD(\chi') \bfth, \chi') \, ,
\end{equation}
\pr{and plays the role of projected density contrast. We will also assume a small angle approximation and consider that the integral on Eq. \eqref{delta_projected} is performed in a cylindrical region instead of a conic one. We refer to \cite{Bernardeau:1994hn, Bernardeau:1996un, bernardeau2000construction} to a detailled discution of the approximations assumed here}.

For reasons that will be explicit in the next paragraph we are led to consider the smoothed convergence \pr{over cylindrical regions with aperture scales $\theta_i$}:
\begin{equation}
\label{kappa_sup}
\kappa_{_<}(\theta_i) := \int_0^{\theta_i} \frac{\dd^2 \bfth}{\pi \theta_i^2} \kappa(\bfth) 
\end{equation}
for a given sequence of scales $\theta_1, \theta_2, \ldots, \theta_N$ of particular interest. We observe that, except for a multiplicative factor, this corresponds to the smoothing of the surface density $\delta_{2D}$. We shall, then, define
\begin{equation}
\label{hkappa}
\hkappa(\theta) = \frac{\kappa_{_<}(\theta)}{ \w(\chi_{_S}) } \, 
\end{equation}
as a normalized convergence. 

\subsection{The rate function of the initial field configuration}

We will consider the evolution of \pr{the effective} 2D density field on the lens plane: as a first step, a Gaussian distributed density field $\tau_{2D}$ is set, and for this field we define $\hkappa^{lin}(\theta)$ as in Eq. \eqref{hkappa}, but using $\tau_{2D}$ in Eq. \eqref{kappa_tvm}. We want to compute the elements $\Sigma_{ij} := \langle \hkappa^{lin}(\theta_i) \hkappa^{lin}(\theta_j) \rangle$, $1 \leq i, j \leq N$ of the covariance matrix $\Sigma$ for this gaussian field. On small-angle approximation one obtains \cite{Bernardeau:1994hn}:

\begin{equation}
\label{Sigma}
\Sigma_{ij} =  \int_0^\infty \frac{\dd k_\perp k_\perp}{2 \pi} P(k_\perp) W(\mD \theta_i k_\perp) W(\mD \theta_j k_\perp)
\end{equation}
where 
\begin{equation}
W(x) = \frac{2 J_1(x)}{x}
\end{equation}
is the Fourier transform of the top-hat filter in two dimensions. We normalize $D_+$ to be unity at current time. We assume here $P(k_\perp) = A k_\perp^n$, $-2 \leq n \leq 1$.

As observed in appendix \ref{Large_dev}, the collection $\{\hkappa^{lin}(\theta_i)\}_{1 \leq i \leq N}$ of correlated gaussian random variables obeys the LDP with rate function:
\begin{equation}
\label{rate_func_I}
I(\hkappa^{lin} (\theta_1), \ldots, \hkappa^{lin} (\theta_N)) = \frac{\sigma^2(\theta_N)}{2} \sum_{i j} \Xi_{ij} \, \hkappa^{lin} (\theta_i) \,  \hkappa^{lin} (\theta_j)
\end{equation}
where $\Xi = \Sigma^{-1}$, and $\sigma^2(\theta_N) = \Sigma_{NN}$. When we take the limit $\sigma^2(\theta_N) \to 0$, the rate function determines the exponential decay rate for the probability density function associated to the random variables.

\subsection{The mapping between the initial configuration and the final configuration}
\label{ldp_top_hat}
We assume now that the $\tau_{2D}$ is a density fluctuation produced by a gas of non-interacting particles obeying continuity, Euler and Poisson equations with azimuthal symmetry, i.e., a cylindrical collapse. If this dynamics can be solved, a map connecting linear and non-linear overdensities can be established. If we consider matter contained in a cylindrical region, Gauss theorem will provide us the relation of the radii given initial and final density. For what concerns this work, it is sufficient to know that the normalized non-linear density can be approximated in terms of the linear density as \cite{Bernardeau:1994hn}:
\begin{equation}
\label{zeta}
\zeta(\tau_{2D}) = \frac{1}{\left(1 - \frac{\tau_{2D}}{\nu} \right)^\nu} \, , \qquad \nu = \frac{\sqrt{13}-1}{2} \, .
\end{equation}

We can therefore construct a new family of random variables to describe the convergence produced by the non-linear evolution of $\tau_{2D}$:

\begin{equation}
\label{kappa_hat}
\hkappa(\vartheta_i) = \zeta( \hkappa^{lin}(\theta_i) ) -1 \, \qquad \vartheta_i = \theta_i/\zeta( \hkappa^{lin}(\theta_i) )^{1/2} \, .
\end{equation}
Since we assume no shell-crossings, the angular scales $\vartheta_i$ are related to the initial scales $\theta_i$ by the constraint of mass conservation inside a given shell. The family of random variables $\{ \hkappa(\vartheta_i) \}_{1 \leq i \leq N}$ is obtained from $\{ \hkappa^{lin}(\vartheta_i) \}_{1 \leq i \leq N}$ by the continuous function $\zeta$, and therefore new family of random variables also obeys the LDP as a consequence of the \emph{contraction principle} in Large Deviations Theory (see appendix \ref{Large_dev}). 

\subsection{The rate function of the final field configuration}

The contraction principle states that the rate function for the new family will be given by (see Eq. \eqref{rates_contraction}):

\begin{equation}
\Psi(\hkappa(\vartheta_1), \ldots, \hkappa(\vartheta_N)) = \inf_{\hkappa^{lin}} I(\hkappa^{lin}(\theta_1), \ldots, \hkappa^{lin}(\theta_N)) \, ,
\end{equation}
where $\inf_{\hkappa^{lin}}$ stands for the infimum taken over the collection $\{\hkappa^{lin}(\theta_i)\}_{(1 \leq i \leq N)}$ such that $\hkappa(\vartheta_i) = \zeta( \hkappa^{lin}(\theta_i) )-1$. In the domain in which $\zeta$ is bounded we can perform the inversion $\hkappa^{lin}(\theta_i) = \zeta^{-1}[1+\hkappa(\vartheta_i)]$, that we may also write as $\hkappa^{lin} ( \hkappa(\vartheta_i))$. We can therefore write,
\begin{eqnarray}
\label{RateFunction}
\Psi(\hkappa(\vartheta_1), & \ldots &, \hkappa(\vartheta_N)) = \frac{\sigma^2(\vartheta_N)}{2} \nonumber\\ & & \hspace{-1cm}\times \sum_{i j} \Xi_{i j} \hkappa^{lin} ( \hkappa (\vartheta_i)) \hkappa^{lin} ( \hkappa (\vartheta_j)) \, .
\end{eqnarray}
Again $\Xi = \Sigma^{-1}$, $\Sigma$ being the matrix whose elements are given in Eq. \eqref{Sigma}. 

The Legendre-Fenchel transform of the rate function is the \emph{scaled cumulant generating function} (SCGF), from which all the cumulants can, in principle, be derived (see appendix \ref{Large_dev}).

\subsection{The single cell case}

\begin{figure}[!ht]
\centering
\includegraphics[width=7.5cm]{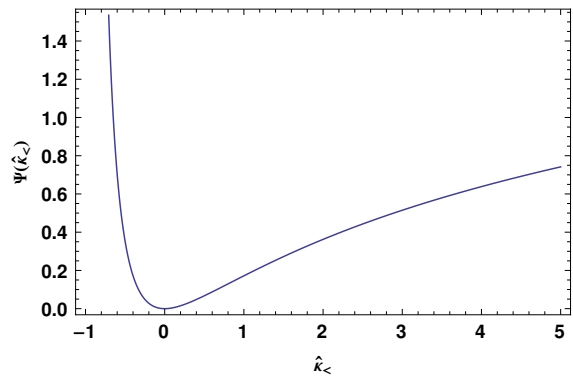}
\caption{The rate function given in Eq. \eqref{rate_function_n1} for $n=-1.5$. The rate function is convex around the origin but changes the convexity for $\hkappa \approx 0.8$.}
\label{rate_func_cyl}
\end{figure}

In order to summarize and illustrate the rate function, SCGF, their relations and role on the derivation of observable quantities, we will consider the convergence filtered at one given scale, i.e., we take $N=1$ in \eqref{RateFunction}. The rate function in this case will be given by
\begin{equation}
\label{rate_function_n1}
\Psi(\hkappa(\vartheta)) = \frac{\sigma^2( \vartheta) \, (\hkappa^{lin}(\hkappa(\vartheta)))^2}{2 \sigma^2( \vartheta \, \zeta^{1/2}(\hkappa^{lin}(\theta)))} \, .
\end{equation}
If $P(k) \propto k^n$, then $\sigma^2(x) \propto x^{-(n+2)}$ in 2D dynamics. We observe from the graph of the function $\Psi(\hkappa)$ shown in Fig. \ref{rate_func_cyl} that this function is not globally convex. Indeed there is a critical value $\hkappa^c$ where there is a change of convexity. For $-2 \leq n \leq 1$, however, $\hkappa^c > 0$ indicating that the rate function is convex in a neighborhood of the origin.

When $\sigma^2 \to 0$, the scaled cumulant generating function is the Legendre-Fenchel transform of the rate function, i.e, 

\begin{equation}
\label{legendre_transf}
\varphi(\lambda) = \sup_{\hkappa} \left[ \lambda \, \hkappa - \Psi(\hkappa) \right] \, .
\end{equation}
The quantities $\varphi$ and $\Psi$ are said to be \emph{convex conjugate}, and Eq. \eqref{legendre_transf} is written simply as $\varphi = \Psi^*$ in some references. If $\Psi$ is globally convex, then $\Psi = \varphi^* = \Psi^{**}$ (i.e., convex conjugation is involutive on the space of convex functions). If $\Psi$ is not globally convex, then $\Psi^{**}$ produces only the convex envelope of $\Phi$. Moreover, $\varphi=\Psi^*$ has points of non-differentiability when $\Psi$ loses convexity. Our rate function, as observed, is not globally convex for all values of the spectral index $n$, what requires careful analysis on the inversion of $\varphi$ to obtain the PDF \cite{bernardeau2014statistics, uhlemann2016back}.

On the strictly convex domain of $\Psi$, however, the Legendre-Fenchel transform reduces to the classical Legendre transform and therefore $\lambda = \frac{\partial \Psi(\hkappa)}{\partial \hkappa}$. If the scaled cumulant generating function is $C^{\infty}$ around the origin, then all the (scaled) cumulants can be obtained from the Taylor expansion of $\varphi(\lambda)$:
\begin{equation}
\label{taylor_cum}
\varphi(\lambda) = \sum_{p=1}^{\infty} \hat{s}_p \frac{\lambda^p}{p!} \, .
\end{equation}
Here $\hat{s}_p = \langle \hkappa^p \rangle_c / \langle \hkappa^2 \rangle_c^{p-1}$. Since $p$-th cumulant is an homogeneous functions of degree $p$, $s_p := \langle \kappa^p \rangle_c / \langle \kappa^2 \rangle_c^{p-1} = \w(\chi_{_S})^{2-p} \hat{s}_p$. For the skewness ($p=3$) we have:
\begin{equation}
\label{s3th}
s_3(\vartheta) = \frac{1}{\w(\chi_{_S})} \left[\frac{36}{7} - \frac{3}{2}(n + 2) \right]
\end{equation}
from where we can depict the inversely proportional dependence of the skewness on $\Omegam$. As it was shown in \cite{Bernardeau:1996un}, $s_3 \approx 42 \, \Omegam^{-0.8}$ for $z_{_S} \approx 1$.

\section{The LDP formulation applied to the observed aperture mass}

We have considered until now a family of random variables produced by filtering the convergence field with a family of top-hat filters associated to scales $\mD \vartheta_1, \ldots, \mD \vartheta_N$ for a given lensing configuration. More general filtering schemes can be proposed, and a very convenient example is the so called \emph{aperture mass}, which is produced by the convolution of the convergence field with a compensated filter $U(x)$, i.e., a filter with the property $\int \dd x \, x \, U(x) = 0$. In order to keep working with normalized quantities, we shall define the \emph{normalized aperture mass} as in \cite{bernardeau2000construction}:

\begin{equation}
\hat{M}_{ap} =  U * \hat{\kappa} \, .
\end{equation}
The scale on which the convergence field is filtered here is determined by a parameter on the definition of $U(x)$. We will compute the aperture mass at the origin here.

\subsection{Filters for convergence and shear}

We should first remark that the aperture mass is defined in terms of a filtering of the convergence field. We would like, instead, to use the top-hat filtered quantity $\hkappa$ on the definition. We remark that the differentiation of Eq. \eqref{kappa_sup} yields $\hat{\kappa}(\vartheta) = \kappa(\vartheta)/w = \hkappa(\vartheta) + \frac{\vartheta}{2} \hkappa'(\vartheta)$. We can thus use integration by parts to write

\begin{equation}
\label{ipp_map}
\Map = \int \dd \vartheta \, \vartheta \, U(\vartheta) \, \hat{\kappa}(\vartheta)
= \int \dd \vartheta  \, \tU (\vartheta) \, \hkappa(\vartheta)
\end{equation}
where $\tU(\vartheta) = -\frac{\vartheta^2}{2} U'(\vartheta)$. The filter $\tU$ satisfies $\int \dd \vartheta \, \tU (\vartheta) =0$. 
Going one step further, we can express the $\hat{M}_{ap}$ in terms an specific filter acting on the shear \cite{schneider1996detection}. For axially symmetrical lenses, we can relate $\hat{\kappa}$, $\hkappa$, and $\hgamma \, (=\gamma/w)$ as:
\begin{equation}
\hgamma(\vartheta)   = \hkappa(\vartheta) - \hat{\kappa}(\vartheta) = - \frac{\vartheta}{2} \hkappa'(\vartheta) \, .
\end{equation}
Assume that there exists a filter $\tQ$ such that
\begin{eqnarray}
\label{gamma_Q}
\Map & = &   \int \dd \vartheta \,  \tQ(\vartheta) \, \hgamma(\vartheta) \nonumber\\ & = & \int \dd \vartheta \, \frac{\tQ(\vartheta) + \vartheta \,  \tQ'(\vartheta)}{2} \,  \hkappa(\vartheta) \, ,
\end{eqnarray}
where the second line was obtained by integration by parts. The relation between the filters $\tU$ and $\tQ$ is, therefore,
\begin{equation}
\tU(\vartheta) = \frac{1}{2} \left( \tQ(\vartheta) + \vartheta \, \tQ'(\vartheta) \right) \, .
\end{equation} 
As before, $\tQ(\vartheta) = -\frac{\vartheta^2}{2} Q'(\vartheta)$. For analytical convenience, we will focus on the filter  $Q(\vartheta) = \e^{- \vartheta^2/
2}$. Other possible compensated filters can be found in the literature \cite{schneider1996detection, bernardeau2000construction, crittenden2002discriminating}.

\subsection{Conveniences and limitations of the aperture mass}

The convergence field can only be obtained through inversion problem and is non-local. Shear field is more directly observed. As discussed in appendix \ref{reduced_shear_app}, observation of ellipticity fields yields a measure of the reduced shear ($g= \gamma/(1-\kappa)$), from which the shear is usually obtained by assuming $\kappa \ll 1$ for weak lensing \cite{kilbinger2015cosmology, munshi2008cosmology}. This leads us to define the \emph{observed aperture mass} as:

\begin{equation}
\label{Map_g}
\Mapg := \int \dd \vartheta \, \hat{g}(\vartheta) \, \tQ(\vartheta) \approx \Map
\end{equation} 
where $\Map$ is given in Eq. \eqref{gamma_Q}. The (\emph{theoretical}) aperture mass $\Map$ has therefore the good analytical property of being expressible in terms of the shear or convergence, but the observable quantity is $\Mapg$. The theoretical and observable aperture masses should agree as long as the convergence is small. Being defined as the filtering of a random field, however, the convergence can exhibit statistical large fluctuations, and therefore the statement $\kappa \ll 1$ is not well defined on itself.

We seek, then, to quantify the role of large deviations on the reconstruction of statistical properties of the aperture mass.

\subsection{General filter and the $\hat{M}_{ap}$ statistics}

We have presented in Sec. \ref{ldp_top_hat} the formalism based on the Large Deviations Theory to derive the SCGF associated to the top-hat filtering of the convergence field for a finite number of filtering scales. A general filtering can be approximated by a weighted composition of top-hat filters as introduced in \cite{Bernardeau:2015khs}, and we will develop this generalization to produce the SCGF for the $\Map$ and $\Mapg$.

We shall assume that the $\Map$ can be approximated by a finite collection of the top-hat filtered convergence $\hkappa(\vartheta_i)$ weighted by the series of coefficients $\tU_i$, $1 \leq i \leq N$ as:

\begin{equation}
\label{map_sum}
\Map \approx \sum_{i=1}^N \, \tU_i(\vartheta_i)  \, \hkappa(\vartheta_i) \, .
\end{equation}
If this is the case, the $\Map$ is a linear combination of the $\hkappa(\vartheta_i)$'s and therefore, by the contraction principle, the rate function for the $\Map$ is given in terms of the initial rate function $I(\theta_1, \ldots, \theta_N)$ through a composition of continuous bounded functions. By the contraction principle (see Eq. \eqref{rates_contraction}),

\begin{eqnarray}
\Psi(\Map) & = & \inf_{\hkappa^{lin}(\theta_i)}  \Bigg[ I(\hkappa^{lin}(\theta_1), \ldots, \hkappa^{lin}(\theta_N)) \nonumber\\ & & + \alpha \left(\Map - \sum_{i=1}^N \tU_i(\vartheta_i) \, \hkappa (\vartheta_i) \right)\Bigg]
\end{eqnarray}
where $\alpha$ is a Lagrange multiplier and $\hkappa(\vartheta_i)$ is given in terms of $\hkappa^{lin}(\theta_i)$ by Eq. \eqref{kappa_hat}.

The weights $\tU_i$ are defined on the image of $\zeta$, i.e., on the filtered non-linear field, but the $\inf$ has to be computed over the initial random variables $\hkappa^{lin}(\theta_1), \ldots, \hkappa^{lin}(\theta_N)$, both being related by the map $\zeta$ defined on Eqs. \eqref{zeta}, \eqref{kappa_hat}.

We assume now that refinements in the partition defined by the filtering scales will eventually lead to the corresponding continuous limit,
\begin{eqnarray}
& & \sum_{i=1}^N \tU_i(\vartheta_i) \, \hkappa (\vartheta_i) \to \int \dd \vartheta \, \tU(\vartheta) \, \zeta(\hkappa^{lin}(\theta))
\nonumber\\ 
&  & =  \int \dd \theta  \frac{ \dd \vark }{\dd \theta} \, \tU [\vark (\theta)] \, \zeta(\hkappa^{lin} (\theta))
\end{eqnarray}
where we stress the dependence of $\vartheta$ on $\hkappa^{lin}(\theta)$  by writing Eq. \eqref{kappa_hat} as:
\begin{equation}
\label{vark}
\vark (\theta) = \theta \, \zeta^{-1/2}(\hkappa^{lin}(\theta)) \, .
\end{equation}

\subsection{The Scaled Cumulant Generating Functions}

Our next goal is write the SCGF on the continuous limit. By the Varadhan's lemma given in Eq. \eqref{varadhan} the scaled cumulant generating function will be the continuous limit of

\begin{equation}
\varphi(\lambda) = \sup_{ {\hkappa^{lin}} } \left[ \lambda \sum_i^N \tU_i(\vartheta_i) \, \hkappa (\vartheta_i) -  I(\hkappa^{lin}(\theta_1), \ldots, \hkappa^{lin}(\theta_N)) \right] \, .
\end{equation}

In order to write the continuous limit of the rate function $ I(\hkappa^{lin}(\theta_1), \ldots, \hkappa^{lin}(\theta_N))$ given in Eq. \eqref{rate_func_I}, we have to give a continuous limit to the matrix $\Xi$. Let us assume, for this sake, the existence of an object $\xi(\theta', \theta'')$ such that
\begin{equation}
\label{invert_xi}
\int \dd \theta' \, \sigma^2(\theta, \theta') \, \xi(\theta', \theta'')  = \Dirac(\theta-\theta'')
\end{equation}
where $\sigma^2(\theta, \theta') $ is given by Eq. \eqref{Sigma} computed over continuous domains.

The desired continuous limit for the SCGF is, therefore,
\begin{eqnarray}
\label{phi_cont}
\varphi(\lambda) & = & \sup_{\hkappa^{lin}} \Bigg[ \lambda \int \dd \theta \, \tU(\vark(\theta)) \, \frac{\dd \vark}{\dd \theta}  \, \zeta(\hkappa^{lin} (\theta)) \nonumber\\ & & - \frac{\sigma^2_{F}}{2} \int \dd \theta \, \dd \theta' \, \hkappa^{lin} (\theta) \, \hkappa^{lin} (\theta') \, \xi(\theta, \theta') \Bigg] \, .
\end{eqnarray} 
Here $\sigma_F^2 = \int \dd \theta \, \dd \theta' \, \sigma^2 (\theta, \theta') \, \tilde{U}(\theta) \, \tilde{U}(\theta')$.  

\pr{As already observed, the normalized aperture mass $\Map$ can be obtained as in Eq. \eqref{ipp_map} through the convolution of $\hkappa$ and $\tU$, or as in Eq. \eqref{gamma_Q} by convolving $\hgamma$ and $\tQ$, and the two different expressions are related by an integration by parts. Integrations by parts can also be applied on Eq. \eqref{phi_cont} to re-express it explicitly in terms of the convergence or shear with corresponding filters. Although equivalent, the different expressions of $\varphi$ allows us perform distinct numerical implementations  and check the performance of the solutions. A second need for a rewriting of Eq. \eqref{phi_cont} in terms of $\hgamma$ and $\tQ$ is to extend the SCGF for $\Map$ into the SCGF of $\Mapg$. We will call the different arrangement of variables \emph{representations} of the SCGF.}

\subsubsection{SCGF on the convergence representation}

The SCGF for $\Map$ in Eq. \eqref{phi_cont} is already given in terms of $\hkappa$ and $\tU$, but it can be expressed in a more numerically suitable form.  Let the filter $\tV(x)$ be defined by:
\begin{equation}
\label{V_def}
\tV(x) = \int_0^{x} \dd y \, \tU(y) \, .
\end{equation}
It follows by integration by parts that:
\begin{eqnarray}
\label{ipp_V}
& & \int_0^{\infty} \dd \theta \, \left( \tU(\vark(\theta)) \, \, \frac{\dd \vark}{\dd \theta} \right) \zeta(\hkappa^{lin} (\theta))  \nonumber\\ & & = - \int_0^{\infty} \tV(\vark(\theta)) \frac{\partial\zeta(\hkappa^{lin}(\theta))}{\partial \hkappa^{lin}} \ \hkappa^{lin'}(\theta)
\end{eqnarray}
as long as the surface term $\tV(\vark) \zeta(\hkappa^{lin}(\theta))|_0^{\infty} = 0$, what is the case if $U(x)$ is a compensated filter. The SCGF on the \emph{convergence representation} is therefore:

\begin{eqnarray}
\label{phi_cont_k}
\varphik(\lambda) & = & - \inf_{\hkappa^{lin}} \Bigg[  \lambda \int \dd \theta \, \tV(\vark(\theta')) \, \frac{\partial\zeta(\hkappa^{lin}(\theta'))}{\partial \hkappa^{lin}}  \, \hkappa^{lin'}(\theta')  \nonumber\\ & & + \frac{\sigma^2_{F}}{2} \int \dd \theta \, \dd \theta' \, \hkappa^{lin} (\theta) \, \hkappa^{lin} (\theta') \, \xi(\theta, \theta') \Bigg] \, .
\end{eqnarray}

\subsubsection{SCGF on the shear representation}

As a direct consequence of Eq. \eqref{gamma_Q} applied to Eq. \eqref{phi_cont}, the SCGF on the \emph{shear representation} is given by:

\begin{eqnarray}
\label{phi_cont_shear}
\varphig(\lambda) & = &  - \inf_{\hkappa^{lin}} \Bigg[  \lambda \int \dd \theta \, \tQ (\vark(\theta)) \,  \frac{ \vark}{2} \, \frac{\dd \zeta(\hkappa^{lin} (\theta))}{\dd \theta} \nonumber\\ & & + \frac{\sigma^2_{F}}{2} \int \dd \theta \, \dd \theta' \, \hkappa^{lin} (\theta) \, \hkappa^{lin} (\theta') \, \xi(\theta, \theta') \Bigg] \, .
\end{eqnarray}
We recall that $ \hgamma  = -\frac{ \vark}{2} \, \frac{\dd \zeta(\hkappa^{lin} (\theta))}{\dd \vark}$.

\subsubsection{SCGF on the reduced shear representation}

We want now to extend the SCGF $\varphig(\lambda)$ to the SCGF for the observed aperture mass $\Mapg$ given in Eq. \eqref{Map_g}. For this sake, we remember that the (normalized) reduced shear can be expressed as:

\begin{eqnarray}
& & \hat{g}(\hkappa^{lin}(\theta), \w)  =  \frac{\hat{\gamma}}{1-\kappa} \nonumber\\ &  & \phantom{\hat{g}(\hkappa^{lin}(\theta))} = - \frac{1}{2} \frac{\vartheta \, \hkappa(\vartheta)'}{1 - \w [ \hkappa(\vartheta) -\frac{1}{2} \vartheta \ \hkappa(\vartheta)']} \, 
\end{eqnarray}
in terms of the variables $\vartheta$, $\hkappa(\vartheta)$, and the projection factor $\w$. \pr{The contraction principle allows us to extend the LDP to any continuous functional of the initial profile, and we can invoke it to replace} $\hgamma$ by $\hat{g}$ in Eq. \eqref{phi_cont_shear} and obtain the SCGF on the \emph{reduced shear representation}:

\begin{eqnarray}
\label{phi_cont_g}
\varphir(\lambda) & = &  - \inf_{\hkappa^{lin}} \Bigg[  \lambda \int \dd \theta \, \tQ (\vark(\theta)) \, \hat{g}(\hkappa^{lin}(\theta), \w) \nonumber\\ & & + \frac{\sigma^2_{F}}{2} \int \dd \theta \, \dd \theta' \, \hkappa^{lin} (\theta) \, \hkappa^{lin} (\theta') \, \xi(\theta, \theta') \Bigg] \, .
\end{eqnarray}

\subsubsection{Summary}

The SCGFs $\varphik(\lambda)$ and $\varphig(\lambda)$ are two (equivalent) expressions the SCGF for $\Map$ given respectively in the convergence and shear representations. The SCGF $\varphir(\lambda)$ is the extension of $\varphig(\lambda)$ where the reduced shear is used as central observable quantity, and therefore corresponds to the SCGF for the physical $\Mapg$.

If the $\inf$s in Eqs. \eqref{phi_cont_k}, \eqref{phi_cont_shear}, and \eqref{phi_cont_g} can be computed, the statistical properties of $\Map$ and $\Mapg$ will follow. It should be stressed that the infimum of a functional doesn't need to agree with a minimum of that functional, but we will only focus here on the determination of minima of our functionals of interest.


\section{Practical Implementations}

The actual resolution of the minimization problems obtained in the previous section is not straightforward with no guarantee that it actually converges. We present in the following the different approaches that we effectively employed to solve
the minimization problem and check that we have consistent results. Details on the numerical tests can be found in the appendix \ref{Numericals}.

\subsection{Extremization with the help of the Euler-Lagrange equations}
\label{euler_lagrange_sec}
An extremum of the functional $\varphik(\lambda)$ given in Eq. \eqref{phi_cont_k} can be obtained by imposing
\begin{equation}
\label{first_variation}
\frac{\delta \varphik(\lambda)}{\delta \hkappa^{lin}(\theta)} = 0 \, .
\end{equation}

Indeed, it follows from Eqs. \eqref{ipp_V}, \eqref{phi_cont_k}, and \eqref{first_variation} that the linear profile has to obey the equation:

\begin{eqnarray}
\label{kappa_euler_lagrange}
\hkappa^{lin} (\theta) & = & - \frac{\lambda}{\sigma_{_F}^2} \int \dd \theta' \sigma^2(\theta, \theta') \frac{\delta}{\delta \hkappa^{lin}} \Big[  \tV(\vark(\theta')) \nonumber\\ & & \times \frac{\partial\zeta(\hkappa^{lin}(\theta'))}{\partial \hkappa^{lin}} \ \hkappa^{lin'}(\theta') \Big]
\end{eqnarray}
where
\begin{equation}
\frac{\delta}{\delta \hkappa^{lin}} = \frac{\partial}{\partial \hkappa^{lin}} - \frac{\dd}{\dd \theta} \frac{\partial}{\partial \hkappa^{lin'}} \, 
\end{equation}
is the Euler-Lagrange operator. The term inside squared brackets in Eq. \eqref{kappa_euler_lagrange} has the nice property of being linear on $\hkappa^{lin'}$, what leads to great simplifications. Indeed, the use of the explicit form of $\vark$ given in Eq. \eqref{vark} and the definition of $\tV$ given in Eq. \eqref{V_def} lead directly to:

\begin{eqnarray}
\label{kappa_lin_int}
\hkappa^{lin} (\theta) & = &  \frac{\lambda}{\sigma_{_F}^2} \int \dd \theta' \, \sigma^2(\theta, \theta')  \Big[ \tU(\vark (\theta'))   \nonumber\\ & & \times \, \zeta^{-1/2} (\hkappa^{lin} (\theta')) \,  \frac{\partial \zeta (\hkappa^{lin} (\theta'))}{\partial \hkappa^{lin}} \Big] \, .
\end{eqnarray}

Also from Eq. \eqref{first_variation} it follows that:
\begin{eqnarray}
& & \frac{\sigma^2_{F}}{2} \int \dd \theta \, \dd \theta' \, \hkappa^{lin} (\theta) \, \hkappa^{lin} (\theta') \, \xi(\theta, \theta') = \frac{\lambda}{2} \int \dd \theta \dd \theta'  \hkappa^{lin} (\theta) \nonumber\\ & &  \phantom{aaa} \times  \Big[\tU(\vark (\theta'))\zeta^{-1/2} (\hkappa^{lin} (\theta')) \,  \frac{\partial \zeta (\hkappa^{lin} (\theta'))}{\partial \hkappa^{lin}} \Big] 
\end{eqnarray}
showing that the actual knowledge of $\xi(\theta, \theta')$ assumed in Eq. \eqref{invert_xi} is not necessary to the solution of the extremization problem.

We obtain, finally, 
\begin{eqnarray}
\label{phi_eom}
\varphik(\lambda) & = & - \lambda \int \dd \theta \, \tV(\vark (\theta; \lambda)) \, \frac{\partial \zeta(\hkappa^{lin} (\theta; \lambda))}{\partial \hkappa^{lin}} \, \hkappa^{lin'}(\theta; \lambda) \nonumber\\ & & - \frac{\lambda}{2} \int \dd \theta \, \dd \theta' \, \hkappa^{lin} (\theta; \lambda) \, \Big[ \tU(\vark (\theta'; \lambda))  \nonumber\\ & & \times \, \zeta^{-1/2} (\hkappa^{lin} (\theta'; \lambda)) \, \frac{\partial \zeta (\hkappa^{lin} (\theta'; \lambda))}{\partial \hkappa^{lin}} \Big]
\end{eqnarray}
where $\hkappa^{lin}(\theta; \lambda)$ is a solution of the integral equation \eqref{kappa_lin_int} for each value of $\lambda$, and $\vark (\theta; \lambda) = \theta \, \zeta^{-1/2}(\hkappa^{lin}(\theta; \lambda))$. 

\pr{From one side, the linearity of our action in $\hkappa^{lin'}$ allows to obtain the simple expression for $\hkappa^{lin} (\theta)$ given in Eq. \eqref{kappa_lin_int}. From the other side, the same linearity of the action in $\hkappa^{lin'}$ forbids us of exploring the stability of the extremum through second variations of the action. We will assume that $\hkappa^{lin} (\theta)$ conducts to a maximum of the SCGF on the interval on which it is defined}.

The SCGF is given by Eq. \eqref{phi_eom} as long as Eq. \eqref{kappa_lin_int} can be solved uniquely for each value of $\lambda$. In order to study the existence of unique solutions for this equation, we take the first variation of Eq. \eqref{kappa_lin_int}, obtaining:

\begin{eqnarray}
\label{critical_det}
& & \frac{\sigma^2(\theta, \theta')}{\sigma_{_F}^2} \frac{\delta}{\delta \hkappa^{lin}(\theta)} \Bigg[ \tU(\vark (\theta')) \nonumber\\ & & \times \, \zeta^{-1/2} (\hkappa^{lin} (\theta')) \,  \frac{\partial \zeta (\hkappa^{lin} (\theta'))}{\partial \hkappa^{lin}} \Bigg] = \frac{\Dirac(\theta - \theta')}{\lambda} \, 
\end{eqnarray}
from which we can determine critical values of $\lambda$ on the solution of Eq. \eqref{kappa_lin_int}.

\subsection{Skewness}
Expanding $\varphi(\lambda)$ given in Eq. \eqref{phi_cont} around $\lambda=0$ leads to the expression of cumulants. For the skewness we have \cite{Bernardeau:2015khs}:

\begin{equation}
\label{skew}
\hat{s}_3^\kappa = 3 \nu_2 \frac{ \int \dd x \, \tU(x) \, \Sigma^2(x)}{[\int \dd x \, \tU(x) \, \Sigma(x)]^2} + 3 \frac{ \int \dd x \, x \, \tU(x) \, \Sigma(x) \, \Sigma'(x)}{[\int \dd x \, \tU(x) \,  \Sigma(x)]^2}
\end{equation}
with
\begin{equation}
\Sigma(x) = \int \dd y \, \sigma^2 (x, y) \, \tU(y) \, .
\end{equation}
(Note that the coefficient in front of the second term in Eq. \eqref{skew} is $6/d$, where $d$ is the dimension of space in which spherical collapse is considered).
It is worth to recall that the reduced skewness of the aperture mass is then
\begin{equation}
s_3^\kappa = \frac{1}{\w} \hat{s}_{3}^\kappa
\end{equation}

We can do the same exercise by expanding $\varphir(\lambda)$ around $\lambda=0$. At lowest order, the correction on the skewness introduced by the use of the reduced shear is:
\begin{equation}
\label{corrective_term}
\hat{s}_3^g - \hat{s}_3^\kappa = 6 \, \w \, \frac{\int \dd x \, \tQ(x) \, \frac{x}{2} \, \Sigma'(x) \left( \Sigma(x) + \frac{x}{2} \, \Sigma'(x) \right)}{[\int \dd x \, \tU(x) \,  \Sigma(x)]^2} \, 
\end{equation}
and
\begin{equation}
{s}_3^g - {s}_3^\kappa=\frac{1}{\w}\left(\hat{s}_3^g - \hat{s}_3^\kappa \right)\,.
\end{equation}

\begin{figure}[!ht]
\centering
\includegraphics[width=7.5cm]{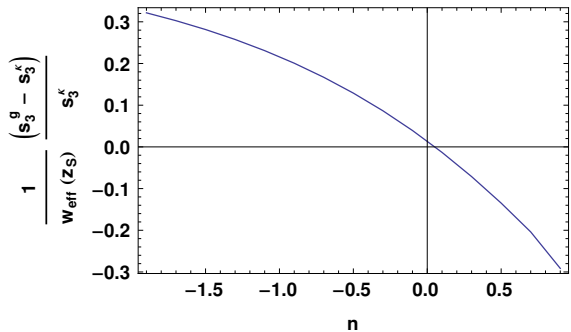}
\caption{Deviation of the skewness for $\Mapg$ and $\Map$ normalized by $\w$. Since $0 \leq \w \leq 0.3$ in Einstein-de Sitter for $z_{_S}=1$, we see that the highest possible correction to the skewness in this case is of order of $10\%$.}
\label{skew_th}
\end{figure}

As expected the expression of $s_{3}^\kappa$ and ${s}_3^g$ depend on the choice of power spectrum and for a sake of simplicity we illustrate our results for power law spectra. More specifically we can analyse the relative importance of the correction term given in Eq. \eqref{corrective_term} by computing the ratio $({s}_3^g - {s}_3^\kappa)/(\w {s}_3)$ which is then independent of $\w$. As shown in Fig. \ref{skew_th}, the impact on the skewness of the distribution of $\Mapg$ has a few percent deviation from the skewness for $\Map$ as function of the spectral index $n$.

\subsection{The SCGF and the rate functions}
\pr{The Euler-Lagrange equations presented in Sec. \ref{euler_lagrange_sec} can be used to obtain $\varphik$ but the action becomes considerably more involved when we are interested in $\varphir$. We treated the problem numerically in two alternative ways: the implementation of Eqs. \eqref{kappa_lin_int} and \eqref{phi_eom}, or the direct extremization methods available in Mathematica \cite{math} such as the FindMinimum routine. As shown in appendix \ref{Numericals}, direct extremization and the Euler-Lagrange methods agree to high precision on the range of $\lambda$ of our interest and for the discretizations of the interval considered.}

\begin{figure}[!ht]
\centering
\includegraphics[width=7.5cm]{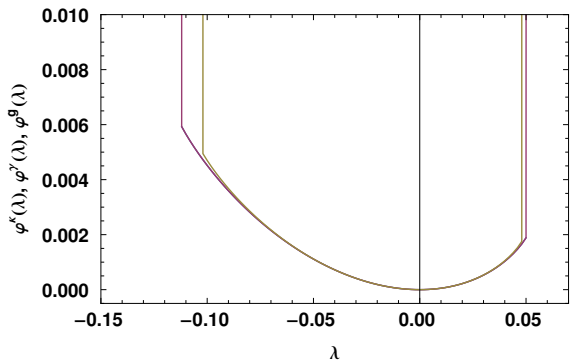}
\caption{The SCGFs $\varphik(\lambda)$, $\varphig(\lambda)$, and $\varphir(\lambda)$ for $n=-1.5$. The curves for $\varphik(\lambda)$ and $\varphig(\lambda)$ coincide as they should. The SCGF for $\Mapg$ is however slightly different and has different critical points. We use $\w=0.1$ here.}
\label{three_phis}
\end{figure}

The general output from the numerical solutions can be seen in Fig. \ref{three_phis}. On this plot $\varphik(\lambda)$ is computed using the Euler-Lagrange strategy and $\varphig(\lambda)$ and $\varphir(\lambda)$ derive from direct extremization. We should remark that although displayed in two different representations we have $\varphik(\lambda) = \varphig(\lambda)$, and this is recovered in the numerical reconstructions (see Sec. \ref{comparison} for more details). We note that moving from the shear to the reduced shear induces some changes on the SCGF and also moves the critical points closer to the origin. The critical points for $\varphik$ can be also obtained from Eq. \eqref{critical_det}.

The SCGF is always a convex function, the same is not true for the rate function. As already noted, and discussed in appendix \ref{Large_dev}, SCGF and rate function can be obtained from each other by Legendre transform as long as the rate function is convex. If the rate function ceases to be convex, the SCGF will still be given by the Legendre-Fenchel transform of the rate function, but will present points of non-differentiability. We then note that the Legendre-Fenchel transform of such a SCGF will produce the convex envelope of the rate function only, and not the rate function itself\footnote{A duality property connects features of the SCGF and the convex envelope of the rate function \cite{2009PhR...478....1T}: if the SCGF goes to infinity at critical points, then the convex envelope of the rate function will have segments of constant derivative beyond critical points. The location of the critical points of the SCGF give the inclination of the affine segments of the rate function, while the lateral derivatives of the SCGF at the critical points give the location of its critical points. We can understand from the LDP that linear segments on the rate function implies exponential decay for the probability density function. 
Unless we know the geometry of the space of solutions for the minimization problem, we cannot know the rate function globally for our problem. All we can access from our method is therefore the segment that corresponds to the non-affine segment of the convex envelope just described.}.

\subsection{The One-point PDF}

We complete our investigations by evaluating the impact of those changes on the shape of the one-point PDF of the
$\Map$ values. The calculation is based on the computation of the inverse Laplace transform of the cumulant generating function (see for instance \cite{bernardeau2014statistics} for details). For a sake of simplicity we perform this calculation using an effective form for the SCGF (as in \cite{bernardeau2000construction}). 
Such an effective form is based on an effective vertex generating function $\zeta_{\eff}(\hkappa^{lin})$ satisfying:
\begin{subequations} 
\label{phi_eff_sist}
\begin{align}
\varphi_{\eff}(\lambda) & =  \lambda \, \zeta_{\eff}(\hkappa^{lin}) - \frac{1}{2} \lambda \,  \hkappa^{lin} \, \zeta_{\eff}'(\hkappa^{lin}) \label{phi_eff_sist_a}\\
\hkappa^{lin}&  =  \lambda \, \zeta_{\eff}'(\hkappa^{lin}) \label{phi_eff_sist_b} \, .
\end{align}
\end{subequations}
where $\zeta_{\eff}(\hkappa^{lin})$ is adjusted so that $\varphi_{\eff}(\lambda) $ provides a good fit to the SCGF we computed, in particular reproducing its critical behaviors. In practice one can get a very good fit with a fifth order polynomial for $\zeta_{\eff}$.

\begin{figure}[!ht]
\centering
\includegraphics[width=7.5cm]{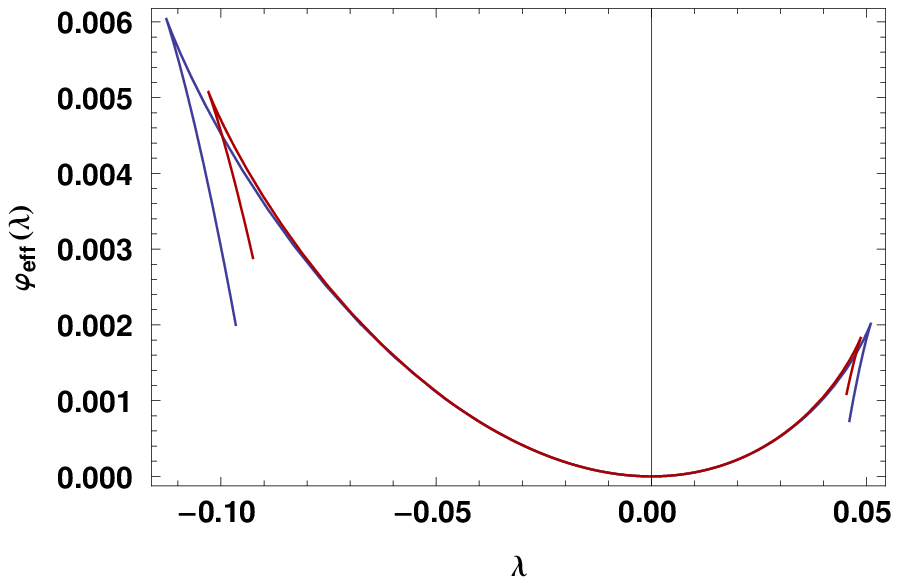}
\caption{The effective cumulant generating functions $\varphik_{\eff}$ and $\varphir_{\eff}$ satisfying Eq. \eqref{phi_eff_sist}. The projection factor $\w=0.1$ is used on the $\varphir_{\eff}$ data.}
\label{phi_eff}
\end{figure}
The effective cumulant generating function obtained this way reproduces extremely accurately the global behavior obtained previously
in particular for its critical points. 

The reconstruction of the one point PDF of $\Map$ is then obtained from the following form,
\begin{equation}
P(\Map) = \int_{- \ii \infty}^{\ii \infty} \frac{\dd \lambda}{2 \pi} \mathrm{exp}[-\lambda \, \hkappa + \varphi_{\hkappa} (\lambda)] \, .
\end{equation}
where the function $\varphi_{\hkappa} ( \lambda)$ is built from the SCGF, 
\begin{equation}
\varphi_{\hkappa} ( \lambda) = \frac{1}{\hsigma^2} \varphik (\lambda \, \hsigma^2)
\end{equation}
in such a way that $\hsigma^2$ matches the expected variance of $\Map$. The actual computation of such inverse Laplace transforms has been described in referenced papers and is based on the integration along the imaginary axis.

\begin{figure}[!ht]
\centering
\includegraphics[width=7.5cm]{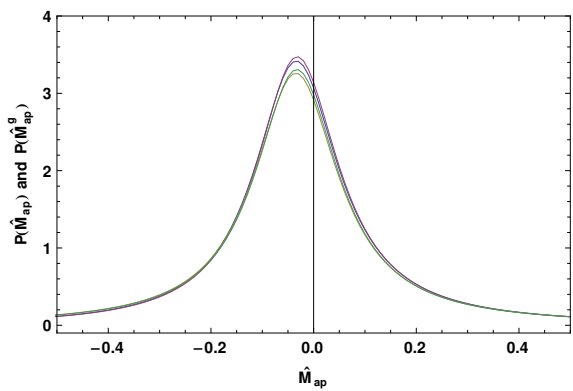}
\caption{PDFs obtained from the inverse Laplace transformation of $\varphik$ and $\varphir$ for $\hsigma=0.4$ (top two curves) and $\hsigma=0.7$ (bottom two curves). The $\Mapg$ is reconstructed for $\w=0.1$. In each case $P(\Map^{g})$ is  sightly larger than $P(\Map)$ for $\Map\approx 0$ exhibiting slightly stronger non-Gaussianities.}
\label{pMap}
\end{figure}

\begin{figure}[!ht]
\centering
\includegraphics[width=7.5cm]{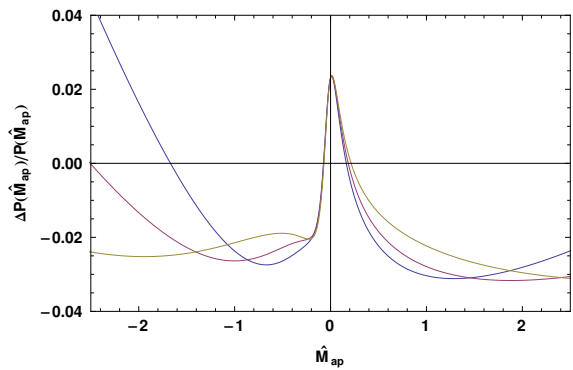}
\caption{Difference of the PDFs shown on Fig. \ref{pMap} for  $\hsigma=0.4$ (blue curve), $\hsigma=0.5$ (red curve) and $\hsigma=0.7$ (yellow curve).}
\label{DiffpMap}
\end{figure}

The resulting PDFs, $P(\Map)$ and $P(\Map^{g})$ built respectively from the shear field and the reduced shear field are shown in Fig. \ref{pMap}. For these ranges of $\Map$ and this value of $\w$ the relative errors are about a few percent as shown on Fig. \ref{DiffpMap}
consistent with our findings concerning the skewness. What these results show however is that the extra non-linearities contained in the reduced shear expression have little effects on the global shape of the PDFs.

\section{Conclusion}

In this paper we present the derivation of the cumulant generating function and the corresponding one-point PDF of  the aperture mass, $\Map$, in a general framework. In particular we take into account that aperture mass can only be built in practice from the measured reduced shear (=shear/(1-convergence)) and will be built in general  from a variety of filter shapes. So, although it is still not a realistic prediction since we did not take into account the exact conical geometry of the observations, our result extends previous computations in two fundamental ways: regarding the shape of the filter and by lifting the usual identification of the shear and the reduced shear.

On a more fundamental level it shows that it is now possible to explore a whole new class of problems, namely the derivation of statistical properties of quantities obtained by general (symmetric) filtering of functionals of the density profiles\footnote{We note that finding a profile satisfying a given condition can be re-expressed in terms of a condition on a non-Markovian random walk. Whether it has a connection to  the computation of halo mass functions in a Press and Schechter like approach is largely an open problem.}. That alone illustrates the reach of the Large Deviation Principle formalism and the fact that it can be exploited in much more elaborate situations than previously thought from direct PT resummations (as pioneered in \cite{1992ApJ...392....1B, 1994A&A...291..697B} and extended in \cite{bernardeau2014statistics}). 

What we found is that the results that were previously obtained, specifically in \cite{bernardeau2000construction}, 
are robust with respect to these extensions. In that study
the compensated filter was indeed assumed to be simply built from the difference of two top-hat filters applied to the convergence field. Our construction is much more elaborate. We find however that the general properties of the cumulant generating functions are left unchanged: they exhibit for instance critical values for both positive and negative values of the parameters whether the $\Map$ is built from the shear of the reduced shear. Differences can be noted however regarding the position of the critical points: as expected the critical values are closer to the origin as large excursions in $\kappa$ lead to stronger non-linear effects. The impact of this effect is however rather mild. We have expressed it in terms of the $\Map$ skewness and the $\Map$ one-point PDF. 

The differences we found are of the few percent range and would require therefore extremely good measurements to be of significant impact.

%
%

Formally the construction of the scaled cumulant generating function is based on the minimization of the a functional form of the density in concentric cells, that is of the whole density profile once we are in the continuous limit. In our study we did not explore in much details how to do such minimization efficiently but checked  that our results were correct from different approaches (that are equivalent in the continuous limit) leaving room for further improvements.


\subsection*{Acknowledgements}
The authors thank C. Pichon and C. Uhlemann for useful comments. This work was supported by the grant ANR-12-BS05-0002 and Labex ILP (reference ANR-10-LABX-63) part of the Idex SUPER of the programme Investissements d'avenir under the reference ANR-11-IDEX-0004-02.
\bibliographystyle{h-physrev}
\bibliography{references}

\appendix
\section{Lensing in Cosmology}
\label{app_lens}
Null vectors in Lorentzian geometry are orthogonal to themselves: let $\gamma$ be a null geodesic, then $\langle \gamma', \gamma' \rangle = 0$ (here $\langle \, \cdot \, , \, \cdot \, \rangle$ means contraction with the spacetime metric, and the prime indicated derivative with respect to the affine parameter). Given $\gamma'$, a four dimension orthogonal basis at each point will be completed by another null vector orthogonal to $\gamma'$ and two space-like vectors. The geometry of a thin light beam will be described by $\gamma'$, that gives its direction of propagation, and by a vector orthogonal to it describing the distance of each null geodesic on the beam. Removing the second null vector by taking the quotient by an equivalence relation \cite{hawking_ellis}, the separation of the geodesics with respect to the fiducial one is described in terms of the so called \emph{Jacobi fields}, i.e., vectors on a two-dimensional space-like screen. The history of the light beam's deformations can be told by the evolution of its sections by successive screens, and the differential equation governing this geometrical evolution is known as \emph{Jacobi equation}:

\begin{equation}
\label{D_pprime}
\mathcal{D}'' + \mathcal{T} \mathcal{D} = 0 \, .
\end{equation}
The object $\mathcal{T}$ is the \emph{optical tidal matrix} that depends on the Ricci and Weyl tensors:
\begin{equation}
\mathcal{T} = \left( \begin{array}{cc}
\mathcal{R} + \mathrm{Re}\mathcal{F} & \mathrm{Im} \mathcal{F} \\
\mathrm{Im} \mathcal{F} & \mathcal{R} - \mathrm{Re} \mathcal{F}
\end{array} \right) \, ,
\end{equation}
where, in components, $\mathcal{R}=\frac{1}{2} R_{\mu \nu} k^{\mu} k^{\nu}$ and $\mathcal{F} = \frac{1}{2} \epsilon^{\alpha} \epsilon^{\beta}  C_{\alpha \mu \beta \nu} k^{\mu} k^{\nu}$  (if $E_1, E_2$ are the orthonormal basis vectors on the space-like screen, then $\epsilon = E_1 + \ii E_2$. Also $k = \gamma'$). The initial conditions to be provided to Eq. \eqref{D_pprime} are $\mathcal{D}(0) = 0$, $\mathcal{D}'(0) = \bold{1}$. The transport problem is solved once one knows $\mathcal{D}$ for all values of the affine parameter \cite{reimberg2013jacobi}.

We want here solve this problem on a perturbed  Friedmann-Lemaitre-Robertson-Walker (FLRW) space-time. Let $\phi$ be the gravitational potential that parametrizes the scalar perturbations on FLRW. If $\phi=0$ the Weyl tensor vanishes and the optical tidal matrix is $\mathcal{T}_0 = K \bold{1}$, where $K$ is the sectional curvature of the space-time. The solution of Eq. \eqref{D_pprime} in this case is $\mathcal{D}(\chi) = \mD(\chi) \bold{1}$, with

\begin{equation}
\label{ang_dist}
\mD(\chi)  = \frac{c/H_0}{\sqrt{1 - \Omegam - \Omegal}} \sinh \left( \sqrt{1 - \Omegam - \Omegal} \frac{H_0 \chi}{c} \right)
\end{equation}
that corresponds to the Jacobi fields in spaces of constant curvature, or the comoving angular diameter as function of the comoving radial distance $\chi$.

To take into account the contributions of the gravitational potential $\phi$ we should go at least to the first order correction on $\mathcal{T}_1$, which includes the Ricci and Weyl tensors associated to the linear perturbations on the metric. We will assume that the lens effect is computed along a null geodesic of the unperturbed space-time, what corresponds to the \emph{Born approximation}.  It can be shown that the first order correction on $\mathcal{D}$ is \cite{bartelmann2010gravitational, peter2013primordial}:

\begin{equation}
\label{D^1/D^0}
\left( \frac{\mathcal{D}_1}{ \mD}\right)_{ab} (\bfth, \chi)  =  - \triangledown_a \triangledown_b \phi_{_L} (\bfth, \chi) 
\end{equation}
where $a, b =1, 2$, and $\phi_{_L}$ is the \emph{lens potential}:
\begin{equation}
\phi_{_L} (\bfth,\chi)  = \frac{2}{c^2} \int_0^{\chi} \dd \chi' \frac{\mD(\chi - \chi')\mD(\chi')}{ \mD(\chi)} \phi (\mD(\chi') \bfth, \chi')  \, .
\end{equation}

The deformation of the beam's sections relative to its spreading on unperturbed spacetime is:
\begin{equation}
\label{deformation}
\frac{\mathcal{D}}{ \mD} = \bold{1} + \frac{\mathcal{D}_1}{ \mD} = \left( \begin{array}{cc} 1 - \kappa - \gamma_1 & -\gamma_2\\ -\gamma_2 & 1 - \kappa + \gamma_1 \end{array} \right)
\end{equation}
where we make a decomposition in trace and traceless contributions by defining \emph{convergence} $\kappa := \frac{1}{2} ( \triangledown_1 \triangledown_1 + \triangledown_2 \triangledown_2) \phi_{_L}$ and \emph{shear} $\boldsymbol{\gamma} := (\gamma_1, \gamma_2)$, with $\gamma_1 = \frac{1}{2} ( \triangledown_1 \triangledown_1 - \triangledown_2 \triangledown_2) \phi_{_L}$, and $\gamma_2 = \triangledown_1 \triangledown_2 \phi_{_L}$.

The determinant of the matrix presented in Eq. \eqref{deformation} vanishes at points corresponding exactly to conjugate points along the fiducial geodesic. Since the presence of conjugate points along the fiducial geodesic is a sufficient condition for multiple imaging \cite{perlick1996criteria}, the vanishing of the determinant roughly indicates the transition from weak to strong lensing. 

\subsection{Lens Map and reduced shear}
\label{reduced_shear_app}

The \emph{lens map} is the map whose differential is the deformation matrix given in Eq. \eqref{deformation}. Instead of writing vectors in $\mathbb{R}^2$, we will use here complex objects. Let $z = x + \ii y \in \mathbb{C}$ and $f(z) = u(x, y) + \ii v(x, y)$ a function of $z$. Using this notation we conclude that:
\begin{equation}
f(z) = u(x, y) + \ii v(x, y) := \left( x - \frac{\partial \phi_L}{\partial x} \right) + \ii \left( y - \frac{\partial \phi_L}{\partial y} \right)
\end{equation}
is the lens map. It follows from the definitions of shear and convergence that:

\begin{subequations}
\label{Beltrami}
\begin{align}
\frac{1}{2} \left[ \left( \frac{\partial u}{\partial x} + \frac{\partial v}{\partial y} \right) + \ii \left( \frac{\partial v}{\partial x} - \frac{\partial u}{\partial y} \right) \right] & =   1 - \kappa \\
\frac{1}{2} \left[ \left( \frac{\partial u}{\partial x} - \frac{\partial v}{\partial y} \right) + \ii \left( \frac{\partial v}{\partial x} + \frac{\partial u}{\partial y} \right) \right] & =   - \gamma
\end{align}
\end{subequations}
where we write $\gamma = \gamma_1 + \ii \gamma_2$. This system of equations is equivalent to Beltrami's equation with Beltrami coefficient $\gamma/(1-\kappa)$ \cite{beltrami_paper}, and the lens map is therefore quasi-conformal. We observe that if $\gamma=0$, then Eqs. \eqref{Beltrami} are the Cauchy-Riemann equations. 

Eqs. \eqref{Beltrami} imply that infinitesimal ellipses are transformed into circles by the lens map. Indeed, 
\begin{equation}
p(x, y) := \frac{1 + \frac{|\gamma|}{1-\kappa}}{1 - \frac{|\gamma|}{1-\kappa}}
\end{equation}
is the ratio of the axes of that ellipse. The quantity 
\begin{equation}
g := \frac{|\gamma|}{1-\kappa}
\end{equation}
defines the \emph{reduced shear}.

The argument of the major axis of the infinitesimal ellipses allows to determine the phase of the Beltrami coefficient \cite{straumann}.

\section{Large Deviation Theory}
\label{Large_dev}
Convergence (or divergence) is one of the most central, and most studied concepts in mathematics. In probability theory different kinds of convergence can be defined. A sequence of random variables $X_n$ converges \emph{strongly} to $X$ if $\mathrm{Pr}[\lim_{n \to \infty} X_n = X] = 1$. Alternatively, a sequence is said to \emph{converge in probability} to a given element $X$ if $\lim_{n \to \infty} \mathrm{Pr}[|X_n - X| \geq \varepsilon] = 0$, for any $\varepsilon > 0$ given. We can define also the \emph{weak convergence} (or convergence in law) by saying that a sequence of probability measures $\alpha_n$ converges weakly to a limiting probability measure $\alpha$ if 
\begin{displaymath}
\lim_{n \to \infty} \int f(x) \dd \alpha_n = \int f(x) \dd \alpha
\end{displaymath}
for every bounded function $f(x)$ on $\mathbb{R}$. Equivalently, if $\phi_n$ and $\phi$ are respectively the characteristic functions of $\alpha_n$ and $\alpha$, $\lim_{n \to \infty} \phi_n(t) = \phi(t)$. The portmanteau theorem in probability theory \cite{varadhan_prob} states that if $\alpha_n$ converges weakly to $\alpha$ on $\mathbb{R}$, and $C \in \mathbb{R}$ is a closed set, then
\begin{displaymath}
\limsup_{n \to \infty} \alpha_n(C) \leq \alpha(C) \, 
\end{displaymath}
while for open sets $G \in \mathbb{R}$,
\begin{displaymath}
\liminf_{n \to \infty} \alpha_n(G) \geq \alpha(G) \, .
\end{displaymath}
If $A \in \mathbb{R}$ is such that $\alpha (\bar{A} - A^o) = 0$ ($\bar{A}$ is the closure of $A$ and $A^o$ its interior), then
\begin{displaymath}
\lim_{n \to \infty} \alpha_n(A) = \alpha(A) \, .
\end{displaymath}

We may think of a class of problems for which a typical value exists, and events far from this typical value will be classified as \emph{rare}. This would be the case for a process described by a probability density function with exponentially decaying tails. Events on the tails are the prototype of rare events. The definition of \emph{large deviation principle} builds on the portmanteau theorem to include the idea that probability measure associated to rare events are exponentially suppressed, and introduces the \emph{rate function} that modulates the exponential decay.

We say that $\{P_{\varepsilon}\}$ obeys the Large Deviation Principle (LDP) with a rate function $I$ if there exists a function $I(\, \cdot \, ) : \mathbb{R} \to [0, \infty]$ lower semicontinuous with compact level sets such that:
\begin{itemize}
\item[i)] For each closed set $C \in \mathbb{R}$
\begin{displaymath}
\limsup_{\varepsilon \to 0} \varepsilon \log P_{\varepsilon} (C) \leq - \inf_{x \in C} I(x)
\end{displaymath}
\item[ii)] For each open set $G \in \mathbb{R}$
\begin{displaymath}
\liminf_{\varepsilon \to 0} \varepsilon \log P_{\varepsilon} (G) \geq - \inf_{x \in G} I(x)
\end{displaymath}
\end{itemize}
If $\inf_{x \in A^o} I(x) = \inf_{x \in A} I(x) = \inf_{x \in \bar{A}} I(x)$, then
\begin{equation}
\label{LDP}
\lim_{\varepsilon \to 0} \varepsilon \log P_{\varepsilon} (A) = - \inf_{x \in A} I(x) \, .
\end{equation}
The parameter $\varepsilon$ has to be identified as some limiting parameter on each problem of interest. For collections of identical, identically distributed random variables (i.i.d.), for instance, $\varepsilon = 1/n$. This definition is general enough to allow Large Deviations Theory to be applied to a large variety of problems of different levels of abstraction, and in general we can replace $\mathbb{R}$ for any complete separable metric space ($\mathbb{R}^n$, for instance) on the definition. We can rephrase the LDP in terms of a family of random variables $\{X_i\}$, by writing $\lim_{\varepsilon \to 0} \varepsilon \log P_{\varepsilon} (\{X_i \} \in A) = - \inf_{x \in A} I(x)$, where we concentrate on Eq. \eqref{LDP} because it can be taken as the definition of the Large Deviation Principle for our needs.

We may ask now two questions: $i)$ are there families of random variables that obey the LDP? $ii)$ What are the most immediate consequences of LDP? To answer the first question, we can quote the famous Cramer's theorem \cite{ellis}: Let $\{ X_i \}$ be a sequence of i.i.d. random vectors on $\mathbb{R}^n$, and $S_k/k := \sum_{i=1}^k X_i/k$ its the \emph{sample mean}. The sequence of sample means $S_k/k$ satisfies the LDP with rate function $I(x) = \sup_{\lambda \in \mathbb{R}^n} [\lambda x - c(\lambda)]$, where $c(\lambda) = \log \mathbb{E}[ \e^{\lambda x}]$ is the cumulant generating function. To have an idea of the origin of the theorem, take $x > \mathbb{E}[X_1]$, $\lambda > 0$,
\begin{eqnarray}
\mathrm{Pr}(S_k/k \in [x, \infty)) & = & \mathrm{Pr}(\lambda S_k \geq k \lambda x)  \nonumber\\
& \leq & \mathrm{exp}[-k \lambda x] \mathbb{E}[ \mathrm{e}^{\lambda S_k}] \nonumber\\
& = & \mathrm{exp}[-k \lambda x] \prod_{i=1}^k \mathbb{E}[ \mathrm{e}^{\lambda X_i}] \nonumber\\
& = & \mathrm{exp}[-k \lambda x] \left(\mathbb{E}[ \mathrm{e}^{\lambda X_1}] \right)^k \nonumber\\
& = &  \mathrm{exp}[-k \lambda x + k \log \mathbb{E}[\mathrm{e}^{\lambda X_1}] ] \nonumber\\
& =& \mathrm{exp}[-k( \lambda x - c(\lambda))]
\end{eqnarray}
Where we use Chebyshev's inequality, independence of $X_i$ and the fact that they are identically distributed to derive this upper bound. This argument does not demonstrate the theorem, but gives a simple illustration of its origins. It points out that rate function and cumulant generating function are related by Legendre-Fenchel transform.

The requirement of independence can be weakened for gaussian random variables. Indeed, for a vector $x$ with mean $\mu$ and covariance matrix $\Sigma$, the rate function is
\begin{equation}
\label{rate_gauss}
I(x) = \frac{1}{2} \sum_{ij} (x - \mu)_i \, \Xi_{ij} \, (x - \mu)_j
\end{equation}
where we denote $\Xi = \Sigma^{-1}$.

Going back to the second question, we will list two consequences of the LDP: the \emph{contraction principle} and the \emph{Varadhan's lemma}. The contraction principle affirms that the image under a continuous map $F$ of families of random variables satisfying the LDP will also satisfy the LDP with rate function:
\begin{equation}
\label{rates_contraction}
J(y) = \inf_{x:F(x) = y} I(x) \, .
\end{equation}
If $F$ is not injective, the contraction principle encodes the  idea that ``any large deviation is done in the least unlikely of all unlikely ways" \cite{der_hollander}. If $F$ is a bijection, on the other hand, $J(y) = I(F^{-1}(y))$. The direct verification of the definition of LDP for a given sequence of probability measures (or random variables) may be prohibitively complicated, and therefore the contraction principle is a powerful tool.

The second important consequence of the LDP that we want to list is the Varadhan's lemma: Consider a family of random variables $\{ X_i^{\varepsilon} \}$, $i=1, \ldots, n$ as a vector in $\mathbb{R}^n$ whose components are $X_i^{\varepsilon}$. If this family of random variables satisfy the LDP with rate function $I(\, \cdot \, )$, and $F: \mathbb{R}^n \to \mathbb{R}$ is bounded, then

\begin{eqnarray}
\label{varadhan}
\lim_{\varepsilon \to 0} \varepsilon \log  \mathbb{E} \left[ \exp \left( \frac{F(X^{\varepsilon})}{\varepsilon} \right) \right]  & = & \sup_{x \in \mathbb{R}^n}[F(x) - I(x)] \, .
\end{eqnarray}
To have an idea of why is this so, we should remember that, because of the LDP, $\mathbb{E} \left[ \exp \left( \frac{F(X^{\varepsilon})}{\varepsilon} \right) \right] \approx \exp \left[ \sup_x [F(x) - I(x)]/\varepsilon \right]$, and use the Laplace method to give an approximative answer to the integral. At the limit $\varepsilon \to 0$ we have the equality. If we consider the simple case where $F(X^\varepsilon) = \sum_i \lambda_i X_i^\varepsilon$, (i.e., the scalar product of $X^\varepsilon$ with a vector $\lambda \in \mathbb{R}^n$) we recognize
\begin{equation}
\label{SCGF}
\lim_{\varepsilon \to 0} \varepsilon \log  \mathbb{E} \left[ \exp \left( \frac{\sum_i \lambda_i  X_i^\varepsilon}{\varepsilon} \right) \right] =: \varphi(\lambda) = \sup_{x \in \mathbb{R}^n}[ \lambda x - I(x) ] \, 
\end{equation}
that is the \emph{scaled cumulant generating function} (SCGF). All the (scaled) cumulants can be obtained from $\varphi(\lambda)$ by partial differentiation. 

The Eq. \eqref{SCGF} also states that the SCGF and the rate function are convex conjugates of each other, or that are related by a \emph{Legendre-Fenchel} transformation. The Legendre-Fenchel transformation reduces to the classical Legendre transformation when the supremum is realized on a maximum of the function under consideration. On the set of convex functions the convex transformation is an involution, and therefore not only the SCGF can be obtained from the rate function by convex transformation, but also the rate function can be obtained from the SCGF. If the rate function is not globally convex, the SCGF is still the Legendre-Fenchel transform of the rate function, but only the convex envelope of the rate function can obtained from the SCGF. Indeed, points of non-differentiability of the SCGF will be associated to points where the rate function loses convexity.

\section{Numerical evaluations}
\label{Numericals}
Our search for the scaled cumulant generating functions for the aperture mass and physical aperture mass conducted us to $\varphik$, $\varphig$, and $\varphir$ given in Eqs. \eqref{phi_cont_k}, \eqref{phi_cont_shear}, \eqref{phi_cont_g} respectively. We must now obtain solutions for the minimization problems.

One possible strategy to approach the solution of the minimization problem is presented in Sec. \ref{euler_lagrange_sec}. The relative simplicity of the functional conducts to $\varphik(\lambda)$ in Eq. \eqref{phi_eom}, once on solves Eq. \eqref{kappa_lin_int} for each value of $\lambda$ in a range of interest. We will refer to it as \emph{Euler-Lagrange (EL) strategy}.

A second possible path is the direct search for minima using the numerical algorithms such as FindMinimum in Mathematica \cite{math}, that we will call \emph{Direct Extremization (DE) strategy}. 

On both cases we will look for numerical solutions initially close to the linear extrapolation 
$(\hkappa^{lin})_{init} (\theta)  =   \frac{1}{\sigma_{_F}^2} \int \dd \theta' \, \sigma^2(\theta, \theta')  \tU(\hkappa^{lin} (\theta'))$.

We should also remark that the non-linear density given in Eq. \eqref{zeta} is singular at $\tau_{2D} = \nu$. The numerical search for solutions may not find good converging paths because of this divergence and in order to circumvent this limitation, we propose a regularization to the non-linear density:
\begin{equation}
\label{zeta_reg}
\zeta^{reg}(\tau_{2D}) = \frac{(1 + \epsilon)^{\nu/2}}{\left(1 - \left(\frac{\tau_{2D}}{\nu} \right)^2 + \epsilon \right)^{\nu/2}}
\end{equation}
where $\epsilon$ is a small parameter (we choose $\epsilon=0.00001$). 

\subsection{Comparison between Euler-Lagrange and Direct Extremization strategies for $\varphik$}
\label{comparison}
We start comparing the performances of the Euler-Lagrange and Direct Extremization strategies for the reconstruction of the same object $\varphik(\lambda)$. 

\subsubsection{The SCGF}

\begin{figure}[!ht]
\centering
\includegraphics[width=7.5cm]{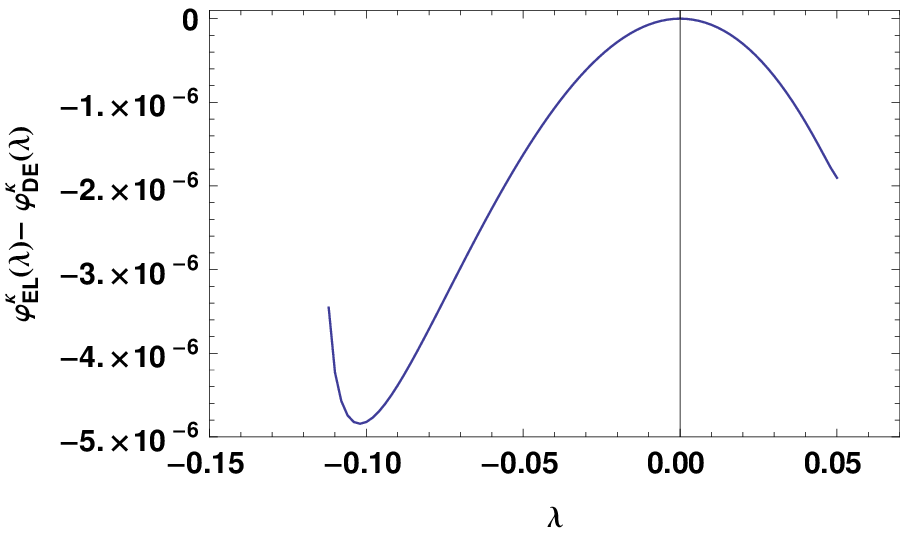}
\caption{$\varphik_{_{EL}}(\lambda) - \varphik_{_{DE}}(\lambda) $for $n=-1.5$.}
\label{diff_phik}
\end{figure}

In order to compare the two numerical strategies on the solution of the equation \eqref{phi_cont_k}, we take the difference of the SCGFs obtained by the EL and DE strategies, as displayed in Fig. \ref{diff_phik}. The small departure of the solutions validate the DE strategy.

\subsubsection{Skewness}

Given the numerical solution $\varphik_{_{EL}}(\lambda)$, we can compute numerically the skewness and compare with the theoretical prediction given in Eq. \eqref{skew}.

\begin{figure}[!ht]
\centering
\includegraphics[width=7.5cm]{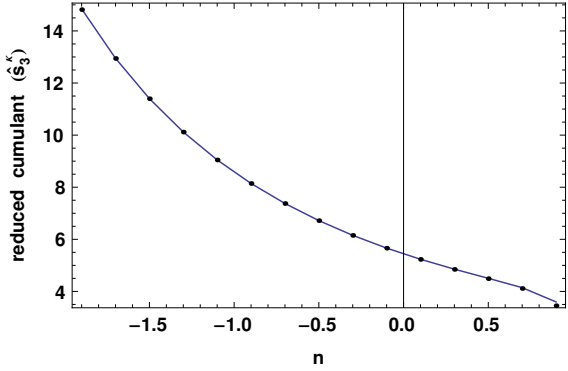}
\caption{Theoretical prediction from Eq. \eqref{skew} and numerical values for the reduced skewness as function of the spectral index.}
\label{cumumalntsEL}
\end{figure}

\begin{figure}[!ht]
\centering
\includegraphics[width=7.5cm]{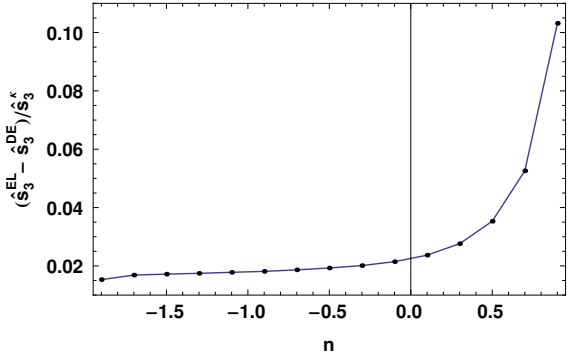}
\caption{The difference of the skweness obteined in EL and DE strategies divided by the theoretical prediction given in Eq. \eqref{skew}. We see that the DE tends to give systematically smaller values for the cumulants for higher values of the spectral index.}
\label{rel_diff_s3}
\end{figure}

We can also compare the performance of the EL and DE strategies on the calculation of the skewness on Fig. \ref{rel_diff_s3}.

The DE strategy has the tendency to copy the initial linear guess on small neighborhoods of the origin, impacting the calculation of the derivatives of the SCGF at the origin. This explains the $1\%$ to $10\%$ error.

\subsection{The skewness for $\Mapg$}
If the use of two different strategies was possible for solving $\varphik(\lambda)$, the same is no longer true if we want to obtain $\varphir(\lambda)$. In this case the functional is no longer linear on $\hkappa^{lin'}$ and the simplifications obtained on the Sec. \ref{euler_lagrange_sec} have no parallel. We are constrained in this case to use direct extremization.

\begin{figure}[!ht]
\centering
\includegraphics[width=7.5cm]{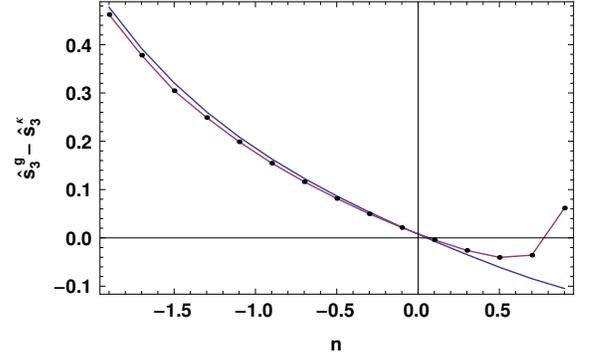}
\caption{Theoretical prediction from Eq. \eqref{corrective_term} and numerical values for the difference of reduced skewness $\hat{s}_3^g - \hat{s}_3^\kappa$ as function of the spectral index for $\w=0.1$. }
\label{corr_term}
\end{figure}

We can verify the numerical values to the theoretical prediction given in Eq. \eqref{corrective_term}. For this sake we compute $\hat{s}_3^g - \hat{s}_3^\kappa$ using the DE strategy for $\w=0.1$. The result is shown in Fig. \ref{corr_term}. As discussed in Fig. \ref{rel_diff_s3}, the DE strategy induces to $1 - 10\%$ error on the skweness, mainly for higher values of $n$, what also impacts the comparison between numerical points and theoretical prediction in Fig. \ref{corr_term}.

 \end{document}